\documentclass{article}

\usepackage{PRIMEarxiv}

\usepackage[utf8]{inputenc} 
\usepackage[T1]{fontenc}    
\usepackage{hyperref}       
\usepackage{url}            
\usepackage{booktabs}       
\usepackage{amsmath}
\usepackage{amsfonts}       
\usepackage{nicefrac}       
\usepackage{microtype}      
\usepackage{lipsum}
\usepackage{fancyhdr}       
\usepackage{graphicx}       
\graphicspath{{media/}}     
\usepackage{soul,color}

\usepackage{xcolor}

\let\olditem\item\renewcommand{\item}[1][black]{\color{#1}\olditem}

\title{Leveraging Deep Learning and Online Source Sentiment for Financial Portfolio Management
}
\author{
  Paraskevi Nousi, Loukia Avramelou, Georgios Rodinos, Maria Tzelepi, Theodoros Manousis, \\ 
  \bf{Konstantinos Tsampazis, Kyriakos Stefanidis, Dimitris Spanos, Manos Kirtas}, \\ 
  \bf{Pavlos Tosidis, Avraam Tsantekidis, Nikolaos Passalis and Anastasios Tefas} \\
  Dept. of Informatics \\
  Aristotle University of Thessaloniki \\
  Thessaloniki, Greece\\
  \texttt{\{paranous,avramell,grodinos,mtzelepi,tmanousis\}@csd.auth.gr},\\ 
  \texttt{tsampazk@gmail.com, kystefan@protonmail.com}, \texttt{\{dvspanos,eakirtas\}@csd.auth.gr},\\
  \texttt{\{ptosidis,avraamt,passalis,tefas\}@csd.auth.gr} \\
}

\begin{document}
\maketitle

\begin{abstract}
Financial portfolio management describes the task of distributing funds and conducting trading operations on a set of financial assets, such as stocks, index funds, foreign exchange or cryptocurrencies, aiming to maximize the profit while minimizing the loss incurred by said operations. Deep Learning (DL) methods have been consistently excelling at various tasks and automated financial trading is one of the most complex one of those. This paper aims to provide insight into various DL methods for financial trading, under both the supervised and reinforcement learning schemes. At the same time, taking into consideration sentiment information regarding the traded assets, we discuss and demonstrate their usefulness through corresponding research studies. Finally, we discuss commonly found problems in training such financial agents and equip the reader with the necessary knowledge to avoid these problems and apply the discussed methods in practice.
\end{abstract}


\section{Introduction}

Financial markets analysis has been and remains a topic of intense research interest since the seminal work of Markowitz \cite{markowitz1991foundations} detailing his theory on portfolio choice, for which he was awarded the Nobel Prize in 1990. The rapid advancements of Machine Learning (ML) and, more specifically those made in the field of Deep Learning (DL) and Deep Reinforcement Learning (DRL), further fueled interest in the field. Financial markets analysts began using ML-based techniques and combining them with their own knowledge of the field \cite{lee1990k}. As early as 1992, Neural Networks (NNs) were already being used for equity index futures trading \cite{trippi1992trading}.

More recently, DL research in financial market analysis has focused on high frequency trading, i.e., an algorithmic financial trading method where high speeds and large volumes are the main characteristics. The kind of data used in works that focus on this type of trading include Limit Order Book (LOB) data \cite{nousi2019machine} as well as candle data for assets such as FOREX or Cryptocurrencies \cite{ahmed2020flf}. Candle data contain the Open, High, Low and Close prices for assets in a requested frequency, e.g., at the minute or hour level. 

Price forecasting is a first step towards solving the very complex task of portfolio management, and has proved to be a sufficiently difficult problem to tackle itself. One way to sufficiently solve it is by transforming the problem into one of classification, i.e., predicting the price \emph{movement} instead of its actual value in the next step \cite{nousi2019machine}. A direction different from this supervised learning one is to train reinforcement learning agents to act as trading agents, operating in this complex financial environment and conducting trades while aiming to maximize their profits and minimize their losses \cite{zarkias2019deep}. More recently, self-supervised learning techniques have been added to this array of DL-based portfolio management methods, in the form of the popular Transformer architecture \cite{daiya2021stock,yang2023deep}, in combination with the aforementioned techniques.

Looking only at the price data, no matter its form, means looking at the problem in a closed environment which is far from a realistic scenario. In reality, financial markets are both influenced by and in turn influence current affairs in all matters of life. This cycle of influence can be captured by public opinion about financial assets or markets in general and has mostly been ignored or not effectively used in studies so far \cite{passalis2022multisource}. Online sources like financial trading websites and news sites as well as public outlets like social media contain a plethora of opinions that experts could analyze and take into consideration when making decisions. By extension it makes sense for DL-based agents to take into consideration the public or expert \emph{sentiment} around the financial assets it deals with.

In this paper we first review recent literature regarding financial portfolio management using supervised learning and reinforcement learning techniques, while also focusing on methods for extracting and exploiting sentiment information regarding the relevant trading assets. 
We identify several challenges that affect DL models when applied to financial data and discuss methods to alleviate the effects of these challenges on the models. First, financial timeseries can be highly non-stationary, as well as very high-dimensional, and exhibit wide variety and great volatility. Typical normalization methods used when training DL models are ineffective in this case and can severely hinder the models' performance, highlighting the need for adaptive normalization mechanisms \cite{nalmpantis2021deep,passalis2021forecasting}. The high-dimensionality and variety in the timeseries can be handled by convolutional and recurrent networks \cite{tsantekidis2020using}, and also by Bag-of-Features models \cite{passalis2020temporal}. 

Second, another great challenge in dealing with financial data arises from the need of price quantization in the supervised learning case of price trend prediction (i.e., classification). This quantization can lead to noisy labels, and its effects can be mended for example by knowledge distillation methods \cite{floratos2022online,tsantekidis2021transferring}. Reinforcement learning techniques can completely overcome this problem as they do not require label quantization, although they offset this advantage with instability during training \cite{tsantekidis2020price}.  This instability constitutes yet another challenge and has been mended in various ways, including knowledge distillation from multiple teachers \cite{tsantekidis2021diversity} and using various different reward shaping methods aiming to stabilize the training process and the profits gained during the agent's deployment in the market \cite{rodinos2023sharpe}.

Finally, when combining sentiment information with price-related features to improve the trading agents' performance, there are several considerations to be made. The quality of sentiment information sources and DL extractors is a subject worthy of investigation itself \cite{passalis2022multisource}. Then, the usefulness of this information pertaining to the financial trading task is another interesting direction to research \cite{passalis2021learning,panagiotatos2022sentiment}. Because of the wide variety of financial data, as well as the intensity of public sentiment regarding some financial assets like cryptocurrencies, extracting and leveraging fine-grained sentiment information is yet another direction that research has been made in \cite{avramelou2023w}.

This paper aims to discuss the aforementioned challenges and present solutions, following the findings of the DeepFinance\footnote{https://deepfinance.csd.auth.gr} project. 

The rest of this paper is organized as follows. In Section~\ref{sec:review} we present a detailed literature review of DL methods with application to financial data, as well as DL methods for sentiment analysis. In Section~\ref{sec:price} we discuss and present the results of works focusing on DL methods for portfolio management based only on price data. Then, in Section~\ref{sec:both} we discuss methods for combining price data and sentiment data for financial portfolio management using reinforcement learning with a focus on methods for stabilizing the performance of these agents. Finally, in Section~\ref{sec:conclusions} we summarize our findings and conclude this study. 

\section{Literature Review}
\label{sec:review}

We conduct a thorough literature review regarding DL models applied to financial data, detailed in Section~\ref{sec:dl_fin}. Then, in Section~\ref{sec:dl_sen} we review methods for sentiment analysis from large textual sources.

\subsection{Deep Learning for Financial Data}
\label{sec:dl_fin}

In the following paragraphs we scour recent literature regarding deep learning methods applied to financial timeseries, starting with simple, feedfowrard networks and moving on to recurrent networks and transformer networks. Furthermore, we discuss more recent research directions such as generative models, explainable DL, and neuromorphic models. Finally, we present recent works on DRL for financial trading, which is the main focus of our work in the context of the DeepFinance project.

\paragraph{Feedforward Neural Networks}
Neural Networks (NNs) have been used in financial market analysis more or less since their popularization after backpropagation was introduced and adopted as their training method \cite{odom1990neural,lee1990k}. More recently, Multilayer Perceptron (MLP) type models and Convolutional Neural Networks (CNNs) have been utilized in various tasks involving financial data. 
MLPs and CNNs can potentially capture non-linear relationships, local dependencies and analyze patterns and anomalies within various features of financial data. In ~\cite{Namdari2021}, the authors used a feedforward MLP to predict the short-term stock trends, by performing a fundamental and technical analysis, additionally employing additional methods to integrate two primary stock market analyses into a hybrid model, outperforming state-of-the-art models as shown in comparisons. In ~\cite{TSANTEKIDIS2017}, a deep learning methodology using CNNs is proposed, which was employed on the stock market and predicted the price movements of stocks. The input used are large-scale, high-frequency time-series derived from the order book of financial exchanges, and the dataset contained more than 4 million limit order events. The proposed method was compared to other MLP and Support Vector Machine (SVM) methodologies, showing that CNNs are better suited for the specific task.

A more recent work ~\cite{CHEN202167}, claimed that other works focused on individual stock information, ignoring correlations between stocks, that can give more insights in stock trend prediction. The proposed method for stock trend prediction used graph convolutional feature based CNNs (GC-CNN), which can capture both stock market  and individual stock features. Randomly selected stocks were used to show superior performance to earlier methods. In ~\cite{THAKKAR2021114800}, a comprehensive survey of deep neural networks, including CNNs, applied on the stock market is presented, showing how DNNs can exploit the influence of various events, financial news, etc., on the financial market. They showcased various methodologies that include CNNs in several ways and with multiple modifications to better apply them on financial data.

\paragraph{Recurrent Neural Networks}
Recurrent Neural Networks (RNNs), and more specifically Long-Short Term Memory (LSTM) networks \cite{troiano2018replicating}, lend themselves naturally to tasks involving financial timeseries, as they are designed to take into consideration important temporal information that MLPs and CNNs mostly disregard. A common LSTM unit is composed of a cell, an input gate, an output gate and a forget gate. The cell remembers values over arbitrary time intervals, while the three gates regulate the flow of information in and out of the cell. This architecture allows LSTM units to retain temporal information that is important according to the task at hand.

LSTM networks have been extensively studied in terms of their application to financial timeseries \cite{yan2021research,tsantekidis2020price,tsantekidis2023modelling}. In \cite{yan2021research} the efficacy of LSTM networks was studied in comparison to that of traditional feedforward networks as well as RNNs, and it was shown that LSTMs outperform the compared architectures. An extensive study made in \cite{samarawickrama2017recurrent} on selected stocks of the Cololmbo Stock Exchange drew similar conclusions, i.e., found that LSTMs outperform other feedforward and recurrecnt architectures, including RNNs and Gated Recurrent Units (GRUs).  

LSTM units can be combined with CNNs in a single architecture to fully harvest the benefits of both models. In \cite{tsantekidis2020using}, a joint CNN-LSTM architecture is proposed and extensively evaluated against CNN and LSTM architectures. The combined CNN-LSTM architecture consistently achieved better performance in both LOB and FOREX data compared to its counterparts. In \cite{rostamian2022event}, a joint CNN-LSTM architecture is studied and applied on candlestick FOREX data to investigate its performance in comparison to the Directional Change (DC) framework, and the effectiveness of the CNN-LSTM model is experimentally validated. In \cite{kazeminia2023real}, a CNN-LSTM model designed for speed was evaluated on cryptocurrency data, specifically BitCoin data, and was shown to achieve better performance than CNNs, LSTMs and GRUs.

\paragraph{Self-supervised Networks}
Self-supervised learning has been gaining traction recently since the introduction of the Transformer architecture in Natural Language Processing tasks \cite{vaswani2017attention}. Since then, variations of this self-supervised architecture have been applied on various other tasks, including computer vision, energy timeseries forecasting and of course financial data analysis. Some of the advantages that make the performance of Transformers Network architectures exceed RNN and LSTM are that they can capture long-range dependencies in sequential data, the parallelization capability that  makes them computationally efficient, and their scalability that allows them to handle large-scale models with billions of parameters, making them capable of capturing complex patterns and nuances in data. These advantages have contributed to the wide adoption of transformers in various domains, including finance. 

Transformers have proven to be effective models for long-range dependencies and interactions in sequential data, making them suitable for time series modeling. Several implementations have been made for stock market price prediction. According to a recent work, using a Data-axis Transformer with Multi-Level contexts can automatically correlate multiple stocks in order to accurately predict the stock movement  without any prior knowledge \cite{DTML}. Another use of Transformer Networks in stock market price time series suggests that it is capable of improving the accuracy of the expected price by using deep feature extraction from text and multiple attention mechanisms to identify dependencies and gather important data \cite{trans_social}. More specifically the implementation is using some historical stock price data combined with social media representations in order to train a transformer network and predict the stock price.

A crucial stage in portfolio management and hedging is price prediction and besides the application in stock markets, there is a need for cryptocurrency price estimation. Since Transformer Networks have shown that can handle long time series with very accurate results, several implementations have been proposed in the cryptocurrency market. In a recent study \cite{crypto_trans_LSTM}, a Transformer is proposed to predict the price of Ethereum, and Bitcoin in order to compare the results that came from an LSTM neural network, both evaluated on the same dataset. The results from the previous research are suggesting that the Transformer Network predictions come with a smaller RMSE than other compared LSTM methods. Another interesting research is focused on  Ethereum fraud detection \cite{fraud_detection} that is taking advantage of the capability of Transformer to capture dynamic sequential patterns in Ethereum transactions and outperform state-of-the-art methods regarding phishing account detection. 

\paragraph{Generative Deep Learning}
Generative DL models have provided significant improvements in terms of performance in tasks like image and text reconstruction as well as question answering, with ChatGPT \footnote{https://openai.com/blog/chatgpt} being perhaps the most well known model in this category. In financial data, generative models could be used for example to fill in gaps in the price timeseries or to generate realistic synthetic data to be used alongside the real timeseries in a data augmentation fashion. 
A generative model is a DL network generating samples from a learned input distribution. More specifically, such models seek to learn a prior probability distribution $P_\theta (x)$ which is parameterized by the neural network’s weights $\theta$. 

Within the context of financial research, generative models find their usage in producing synthetic datasets for various types of tasks, such as stock movement prediction, portfolio optimization or product recommendations. Although many financial datasets are publicly available, there are cases where data is kept private or publicly open datasets are small \cite{dogariu2022generation}. This is the crux of modern statistical inference, since training good prediction models requires large datasets. It is in this setting, however, that generative models can prove handy. A considerable number of research works in literature employ Generative Adversarial Network (GAN) \cite{goodfellow2014generative} architectures for synthesizing financial data. A GAN is trained via an adversarial learning framework, building a generative model through a zero-sum game between a generator (i.e., a network generating data) and a discriminator (i.e., a network trying to guess whether this data comes from a real distribution or not). 

FIN-GAN proposed by \cite{takahashi2019modeling}, is identical to the original GAN implementation proposed by \cite{goodfellow2014generative}, however the study in \cite{takahashi2019modeling} reveals its inherent ability to generate samples preserving important statistical properties of financial time-series, such as linear unpredictability, the heavy-tailed price return distribution, volatility clustering, leverage effects, the coarse-fine volatility, correlation, and the gain/loss asymmetry. In similar works \cite{dogariu2022generation}, \cite{dogariu2021towards} examine a number of GAN implementations for generating realistic time-series data, while also introducing evaluation metrics for measuring the quality of the artificially generated datasets. In \cite{efimov2020using}, proposed Conditional Deep Regret Analytic Generative Adversarial Networks (CDRAGAN), a conditional version of GAN that employs a regularized variation of the discriminator’s loss in order to avoid sharp gradients during training. 

Except for dataset generation, other applications of GANs can be found in financial research. More specifically, Marti \cite{marti2020corrgan} in his seminal work proposed CORRGAN, a GAN implementation for sampling financial correlation matrices.

\paragraph{Explainable Deep Learning}
Research on explainable DL models has been gaining traction, as a solution to their ``black-box'' nature, i.e., the fact that their results are difficult to interpret by experts in the field. 
Arguably, one of the most critical aspects of using DL models in financial data is trustworthiness through explainability. Although there have been improvements in reproducibility, substantial work remains to be done regarding model explainability \cite{olorunnimbe2023deep}. However, studies show that in the last few years number there has been a dramatic increase in articles involving Explainable Artificial Intelligence (XAI) \cite{weber2023applications}. AI models are distinguished between transparent, which are interpretable without any further additions, and so-called post-hoc explainability, which complements existing AI models to create or improve their interpretability. For the latter, there is model-agnostic explainability on the one side, providing explainability regardless of the model, and model-specific explainability on the other, which improves explainability for distinct AI models \cite{arrieta2020explainable}.

One of the recent applications of using explainable DL models is in Deep Reinforcement Learning. One way of adding explainability to Deep RL methods is through employing an explainable artificial intelligence method, class activation mapping (CAM), to explain the network outputs, which computes an activation map for an asset of interest \cite{shi2021xpm}. Moreover, a linear layer in hindsight has been used as the reference model and the relationship between the reward (namely, the portfolio return) and the input (namely, the features) was found out by using integrated gradients \cite{guan2021explainable}. In a task of risk assessment of cryptocurrency investments a Machine Learning approach can be applied to NNs to make the results explainable, by discovering the most important determinants \cite{li2023investigating}.

\paragraph{Neuromorphic Networks}

The increased demand for computation power that required processing among others and large-scale financial time series along with the concerns raised over the sustainability of DL has fueled the research of energy-efficient and ultra-high-speed DL accelerators. Consequently, specialized hardware accelerators have been developed that increase both the training speed and inference speed, while also reducing power and energy consumption. From Graphics Processing Units and Tensor Processing Units~\cite{jouppi2017datacenter} to advanced neuromorphic hardware architectures~\cite{liang2019memristive, jo2010nanoscale} there is a increasing effort for accelerating matrix-based calculations, which cover a significant fraction of the calculations involved during the training and inference of DL models.

The authors in~\cite{kirtas2022robust} proposed an effective training method that is demonstrated in a challenging forecasting problem that involves high-frequency financial time series using a state-of-the-art recurrent photonic architecture, which naturally fits the requirements of such latency-critical applications. The proposed method computes iteratively the appropriate variance for initializing the weights, taking into account the data distribution and the noise sources that intrinsically exist in the photonic components. More precisely, the authors are motivated by the fact that the network should be initialized in a way that allows the information to flow from the input to the output, avoiding to saturate the neurons. The hypothesis is formulated using an information-theoretic criterion, arguing that the mutual information between each layer of the network and the input must be kept as high as possible during the initialization. To ensure that, authors design and employ an auxiliary task that acts as an efficient proxy to this problem. To this end, they proposed a noise-aware training method that appropriately initializes a NN taking into account both the activation functions and noise sources that exist in the photonic implementation and incorporates the physical limitations (noise, constraints, and regularizers) in the training process. 

The proposed method is evaluated on a large-scale high-frequency limit order book dataset (FI-2010)~\cite{ntakaris2018benchmark}. The FI-2010 dataset contains 4 million time series data for five stocks from Nasqad Nordic market in a time frame of 10 consecutive days. The objective of the forecasting task is to predict the direction of mid-price movement (up, down, or stationary) after 10 days. They use the anchored evaluation setup and preprocessing scheme proposed~\cite{nousi2019machine, passalis2021training} using splits 1 to 5 for all experiments conducted. The proposed method significantly improves the performance of the model in the evaluation phase on different levels of noise and applied photonic activation functions. Since the method applies an activation-agnostic initialization taking into account the data distributions and noise sources, it fits also in other applications where easily saturated activation functions are used. As an example, the authors~\cite{kirtas2022learning} have also demonstrated the use of the aforementioned data-driven initialization method in the optical communication domain, while experimental results are presented in~\cite{roumpos2023high, de2022improving}.

\paragraph{Deep Reinforcement Learning}
Finally, we focus on DRL methods for financial data analysis, which is the main kind of learning style used in most recent works on portfolio management. 
DRL is a fast growing filed in Machine Learning, and is the combination of DL and Reinforcement Learning (RL) algorithms. Unlike supervised learning, where tagged data is provided, or unsupervised Learning, where unlabeled data is provided to train a deep neural network, in RL the learning process happens through trial and error. The task is to train an agent, through interactions with a proper environment, to act accordingly in order to solve a given problem. As the agent we describe the deep neural network that chooses actions within a representation of the given problem, which we call environment, which in turn, responds the consequences of the agent's chosen action back to the agent. The goal of the agent is to learn the optimal behavior, called policy, which maximizes a reward signal. Through many interactions within the environment, the agent will learn the optimal behavior. 

Reinforcement Learning has been used with great success in the past years in the fields of robotics, games, autonomous driving and many more. Portfolio theory \cite{markowitz1991foundations} was the basis for traditional portfolio optimization. Based on the covariance matrix of historical returns, the modern portfolio theory allocates the asset with the best return relative to the risk. However, in recent years, the application of DRL to financial data forecasting and portfolio management has gained considerable attention. 
In \cite{jiang2017deep} a RL framework specially designed for the task of portfolio management is proposed. The core of the framework is the Ensemble of Identical Independent Evaluators (EIIE) topology. An IIE is a neural network, which is responsible to inspect the historical data of an asset and evaluate its potential growth for the immediate future. Each asset in the portfolio is assigned with a target weight from the IIE, which define the market action of the RL agent. An asset with an increased target weight will be bought in with additional amount, and that with decreased weight will be sold. EIIE, in addition to the historic market data, accepts an input of the portfolio weights from previous trading periods, in order to enable the RL agent to consider the effect of transaction cost to its wealth. For this purpose, the authors in this work proposed a Portfolio Vector Memory (PVM), in which portfolio weights
of each period are recorded . The EIIE is trained in an Online Stochastic Batch Learning scheme (OSBL). The reward function of the RL framework is the explicit average of the periodic logarithmic returns. The authors experimented on three different neural network architectures, a CNN, a LSTM and a basic RNN, with the RNN achieving the best results in portfolio management.

Another recent DRL work for portfolio management was proposed in \cite{10.1145/3383455.3422540}, in which the authors propose an ensemble of DRL agents to achieve the best Sharpe ratio in 3 month trading periods. Three different agents are trained to create an ensemble for a robust trading strategy, which are a PPO, an A2C and a DDPG agent. These agents are validated using a 3-month rolling window, and the best Sharpe ration achieving agent is picked for the portfolio management for the next three months. The authors reason behind this choice, is that each of these agents is sensitive to different type of market trends. 

In \cite{Wang_Huang_Tu_Zhang_Xu_2021}, the authors propose DeepTrader, a deep RL method which takes into account the market conditions. The proposed model aims to adjust the allocation between long and short funds based on macro market conditions, in order to reduce the risk associated with market fluctuations. It uses the negative maximum drawdown as a measure of risk. Additionally, the model includes a unit that evaluates individual assets by analyzing historical data and considering the rate at which their prices are rising. The model captures both temporal and spatial dependencies between assets using a specific type of graph structure. The authors found that the estimated causal structure provides the best understanding of how assets are interconnected, compared to industry classification and correlation. The two units work together to generate a well-suited portfolio that aligns with the market trend and effectively balances return and risk.

\subsection{Deep Learning for Sentiment Analysis}
\label{sec:dl_sen}


Sentiment Analysis is a popular Natural Language Processing (NLP) task involving classifying text based on its sentiment. The most common classification labels are positive, negative, and neutral, although there exist additional classifications with more classes that represent a wide range of emotions, including happy, sad, angry, and others. The task of sentiment analysis is approached with several methods. As a first step, lexicon-based approaches, such as VADER \cite{hutto2014vader}, have been developed to classify text based on sentiment. These approaches use the dictionary of words with sentiment scores to indicate their sentiment and determine the sentiment of texts by summing the sentiment scores of all words. In addition, machine learning algorithms have been developed for performing sentiment analysis tasks. These types of methods can be Logistic Regression \cite{ramadhan2017sentiment}, Naive Bayes \cite{dey2016sentiment}, and Support Vector Machines \cite{mullen2004sentiment}. More specifically, in these methods, the texts are transformed into numeric vectors that are used in order to train machine learning models as sentiment analyzers. However, sentiment analysis has also been performed with deep learning neural networks. Different types of deep learning neural network architectures have been used in order to train sentiment analysis models. The state-of-the-art is \emph{Bidirectional Encoder Representation Transformers} (BERT) \cite{devlin2018bert}, which is a Transformer based model. 

BERT is a deep neural network developed by Google that is used in NLP tasks including text classification, machine translation, and question answering, among others. BERT models are state-of-the-art in a variety of NLP tasks. The architecture of BERT is encoder-decoder. In more detail, the BERT model is made up of a number of encoder layers with self-attention heads. On top of that, they use embedding layers to represent text as numeric vectors depending on context. The model is trained on a large text corpus, such as Wikipedia or BookCorpus, using two language modeling tasks: Masked Language Modeling and Next Sentence Prediction. This resulted in the development of a model capable of understanding the context of the text in an effective way. 

BERT classification models have been widely used in the development of sentiment analysis models \cite{xu2019bert,sousa2019bert}. These models consist of a BERT model with a linear classification layer on top that predicts the sentiment labels. As previously stated, the BERT is already pre-trained to understand text, but it cannot perform sentiment analysis without additional training. This is known as finetuning, and it involves training the pre-trained BERT model with a specific dataset containing text and its corresponding label-sentiment. This enables us to develop a model that can predict sentiment in text. 

Sentiment analysis from various sources is intuitively an important aspect of financial markets analysis, as it allows researchers to view the problem in a more spherical view, i.e., as part of its environment, instead of a closed box which is an unrealistic setup. Various research directions can be explored in this case, ranging from learning the importance of different sources, combining multimodal sources and training DL models for sentiment extraction. Furthermore, an explainability angle could be explored to provide explanations for the model's predictions in a format more easily understandable by humans. For financial texts however, finetuning sentiment analysis models on this type of data is crucial as this type of test set can differ substantially from the training set of these models, leading to subpar performance.

FinBERT \cite{araci2019finbert} is an example of finetuning a BERT model to develop a sentiment analyzer. FinBERT is a sentiment analysis model based on BERT that is further trained in a financial corpus that contains financial-related text and then finetuned using a dataset that contains financial-related text along with corresponding labels that describe its sentiment as positive, negative, or neutral. The BERT model must have to be trained further since the financial domain has specialized vocabulary that we want the model to recognize. CryptoBERT \cite{passalis2022multisource} is another example of a fine-tuned BERT model. Analogous to the FinBERT model, CryptoBERT is a BERT model that has been trained in a cryptocurrency-related corpus and then fine-tuned using a dataset including text on cryptocurrencies as well as sentiment labels, specifically positive, negative, and neutral. Similarly, we can train a BERT model using corresponding datasets to develop sentiment analysis models for particular applications.

\section{Background: Deep Learning for Financial Trading}
\label{sec:price}

\subsection{Performance Metrics}
\label{sec:metrics}

Supervised learning criteria such as accuracy, precision, recall, and f1 score can be used to compare various models' performances \cite{nousi2019machine}, the same as in other classification tasks. Of course, comparison between models using classification metrics is only possible for models trained in a supervised manner.

To measure and compare the performance of different models on the same set of assets, the utilization of metrics such as the \emph{Profit and Loss} (PnL) measure can be used. PnL can be calculated through the simulation of the actual profit or loss that a trader would have incurred if they had executed the trades. More specifically, PnL is calculated as:

\begin{equation}
    PnL = \sum_{t=1}^N \delta_t r_t  -  |{ \delta_t - \delta_{t-1} }| c,
\end{equation}
where $N$ denotes the total duration of the back-testing period (number of time-steps),  $r_t$ is the return at time step $t$, $c$ is the commission paid for realizing profits/losses and $\delta_t$ is an index variable used to indicate the current position, which is defined as:
\begin{equation}
    \delta_t=\begin{cases}
            -1, &\text{if agent holds a short position at time-step $t$} \\
            1, &\text{if agent holds a long position at time-step $t$} \\
            0, &\text{if the agent is not in the market at time-step $t$}
            \end{cases}. 
\end{equation}

The \emph{Sharpe ratio} \cite{SharpeRatio} is a measure used to compare the return of an investment with its associated risk and provides an insight that returns over a period of time may indicate volatility. It is defined as:

\begin{equation} \label{eq:9}
   SR = \frac{R_p-R_f}{\sigma_p}
\end{equation}

where $R_p$ is the return of a portfolio and is calculated as the mean value of returns, normally, monthly returns, $R_f$ is the risk-free rate and refers to the anticipated return that an investor can expect to receive from an investment, assuming that the investment carries no associated risk, and $\sigma_p$ is the standard deviation of the returns. 

In the realm of Deep Reinforcement Learning, the reward function employed often includes the metrics \emph{PnL} and \emph{Sharpe ratio}. These metrics may be utilized either in isolation or in conjunction with one another as it is done in \cite{rodinos2023sharpe}.



Furthermore, we take into consideration the computational cost of the method whenever possible, as time constraints are present when trading in high frequencies.

\subsection{Supervised Learning}

Supervised learning for financial trading involves training a model using historical data to predict future price movements and make trading decisions. In supervised learning, the model learns from labeled examples, where each example consists of input features, such as technical indicators, market data, and a corresponding target label (e.g., whether to buy, sell, or hold a financial instrument)~\cite{nousi2019machine}. More specifically, the ground labels $l_t$ for training the model to forecast price movements at time $t$ is usually defined as:
\begin{equation}
l_t = \begin{cases}
1  &\text{if $\frac{p_{t+1}}{p_t} - 1 > c_{thres}$}\\
-1 &\text{if $\frac{p_{t+1}}{p_t} - 1 < -c_{thres}$}\\
0  & \text{otherwise},
\end{cases}
\end{equation} 
where $p_t$ denotes the close price at time $t$ and  $p_{thres}$ denotes the threshold for considering that a price movement is a potential candidate for performing a profitable trade. Therefore, the label ``1'' corresponds to a long position, the label ``-1'' to a short position, while the label ``0'' indicates market conditions that probably do not allow the specific agent to perform profitable trades, i.e., the agent should exit the market.  

It is crucial to set a threshold for selecting when to buy, sell, or stay out of the market to manage risk and optimize trading performance. This threshold determines the confidence level required for the model to make a trading decision. By setting a threshold, traders can control the level of risk they are willing to take and avoid making trades based on uncertain predictions. For example, a trader may set a higher threshold for buying or selling to reduce false positives and avoid unnecessary transaction costs. Conversely, setting a lower threshold may increase the probability of catching potentially profitable opportunities but could also lead to more false negatives (missed trading opportunities). Finding the right threshold involves a trade-off between risk tolerance and the desire for accurate predictions to optimize trading strategies. Typically, $p_{thres}$ is set to a value high enough to overcome any commission fees, as well as to account for price slippage that might occur.  Please note that during back-testing, the consecutive ``long'' or ``short'' positions do not lead to multiple commissions (since the agent simply keeps the already existing position open), while the exit position (``0'') closes the currently open position and materializes any gain/loss acquired. It should be also noted that when training with such handcrafted labels, instabilities often arise, leading to models with significant deviation when performing back-testing. Such phenomena can be tackled by appropriately using distillation approaches, e.g., either by transferring knowledge by other ensembles of models~\cite{floratos2022online} or even from existing trading strategies~\cite{tsantekidis2021transferring}, allowing for fine-tuning them, while keeping a prior that is known to work well.

Another important issue that often arises when analyzing financial data is non-stationarity and distribution shifts that often occur to the input. Early results demonstrated that identifying the input distribution and then adaptively normalizing the input can lead to improved performance~\cite{passalis2020adaptive}.  Subsequent works demonstrated the potential of such adaptive input normalization approaches~\cite{passalis2019deep}, while further extending it to improve stability~\cite{passalis2021forecasting}, as well as  by adding bi-linear structures~\cite{tran2021data}. 

Further improvement in the performance of such supervised models can be obtained by appropriately incorporating structures that model the behavior of the data. Some early positive results were obtained by using the Bag-of-Features model to model the input time-series~\cite{passalis2017time}, demonstrating better robustness due to the noise-resilient quantization involved in the process. However, at the same time, this approach can discard useful information. To this end, the Bag-of-Features model has been extended to allow for better introducing temporal information by separately modeling the behavior of the input time-series at different horizons~\cite{passalis2018temporal}. At the same time, further improvements have been obtained by switching from traditional Bag-of-Features formulations into logistic ones, that mitigate many optimization issues that often occur~\cite{passalis2019deepa, passalis2020temporal}, such as vanishing gradients. Using attention-augment bilinear networks~\cite{tran2018temporal, shabani2022multi}, tensor representations~\cite{tran2017tensor} and data-driven architecture learning~\cite{tran2019data} can also lead to further improvements, depending on the nature of the input data.

\subsection{Reinforcement Learning}

One of the main disadvantages of using supervised learning for portfolio management is heuristically finding the threshold of price changes under which we can consider the price as unchanged. This not only leads to data imbalance, but also incorporates subjectivity into the model making it harder to compare with models that use different thresholds. Another main disadvantage is the disconnection of these categorical labels describing the price movement direction from the \emph{amount} of the corresponding price change. This means that a model with better accuracy can ultimately lead to worse profits or even losses, as the price change isn't typically incorporated into the learning objective. 

Reinforcement Learning aims to solve these drawbacks by discarding any need for categorization and aiming to improve the PnL metric itself instead of a classification metric without any tie to actual asset prices. The reward at each time step $t$ is typically defined as:
\begin{equation}
\label{eq:reward}
r_t=\frac{p_{t}-p_{t-1}}{p_{t-1}} \cdot a_t - c_t,
\end{equation}
where $p_t$ is the current price of the asset and $a_t$ is the action that the agent has chosen, i.e., $1$ for the long position, $-1$ for the short position and $0$ when the agent has exited the market (so it received no profit or loss). Note that usually exiting the market does not terminate the training process and the agent has the opportunity to reenter in it. The commission the agent pays at each time step is denoted by $c_t$ in (\ref{eq:reward}) and it is calculated as:
\begin{equation}
c_t=|a_t-a_{t-1}| c.
\end{equation}
where $a_t$ and $a_{t-1}$ are the current and previous actions respectively and $c$ is the commission fee. Again, $c$ is typically set to a value slightly higher than the actual commission paid to account for price slippage and possible risks. Note that the difference between the two actions is $0$ when the agent holds the same position, $1$ when the agent switches its position from $1$ or $-1$ to $0$ or vice versa and $2$ when the agent switches from $-1$ to $1$ or vice versa. Therefore, the agent is penalized when it frequently changes its position. 

To further improve the stability of the training process, sometimes, the average price over a future horizon can be used, i.e., 
\begin{equation}
\label{eq:reward}
r_t=\frac{(m_{t}-p_{t-1})}{p_{t-1}} \cdot a_t - c_t,
\end{equation}
where $m_{t}$ is the average of the next ten prices of the asset.  This reward is not equivalent to the actual  PnL, since the average of the next 10 prices is used instead of the immediate next price. However, this modification allows for significantly improving the training stability, by filtering out potential noise~\cite{nalmpantis2021improving}.

One important issue that often arises when training Deep Reinforcement Learning (DRL) agents for trading is instabilities during the training process, which can lead to a vastly different performance of the agent during the backtesting. To address these challenges several methods have been proposed. First, auxiliary objects, such as price trailing, have been proposed~\cite{zarkias2019deep}. Such approaches are compatible with any DRL methodology~\cite{tsantekidis2020price} and gradually guide the agent by employing an easier-to-solve objective, i.e., to simulate ``driving'' across the input time series.  Using distillation can also further improve the performance of the agents~\cite{tsantekidis2020improving}, especially when using pools of diversified agents trained on different currencies~\cite{tsantekidis2021diversity}. Furthermore, adaptive normalization approaches can be also incorporated into DRL agents, similarly to supervised ones, allowing for mitigating distribution shifts~\cite{nalmpantis2021deep}. Finally, note that the agents can be also extended to handle both market and limit orders, when DRL approaches that can handle continuous action spaces are used, leading to superior performance~\cite{tsantekidis2023modelling}, at the cost of higher complexity of modeling limit order book information and incorporating it into the used trading environment.

\section{Adapting to Challenges in DL-based Financial Trading}
\label{sec:both}

\subsection{Addressing the Non-stationarity, High-dimensionality and Volatility of Financial Timeseries}
%

\begin{figure}
    \centering
    \includegraphics[width=0.5\linewidth]{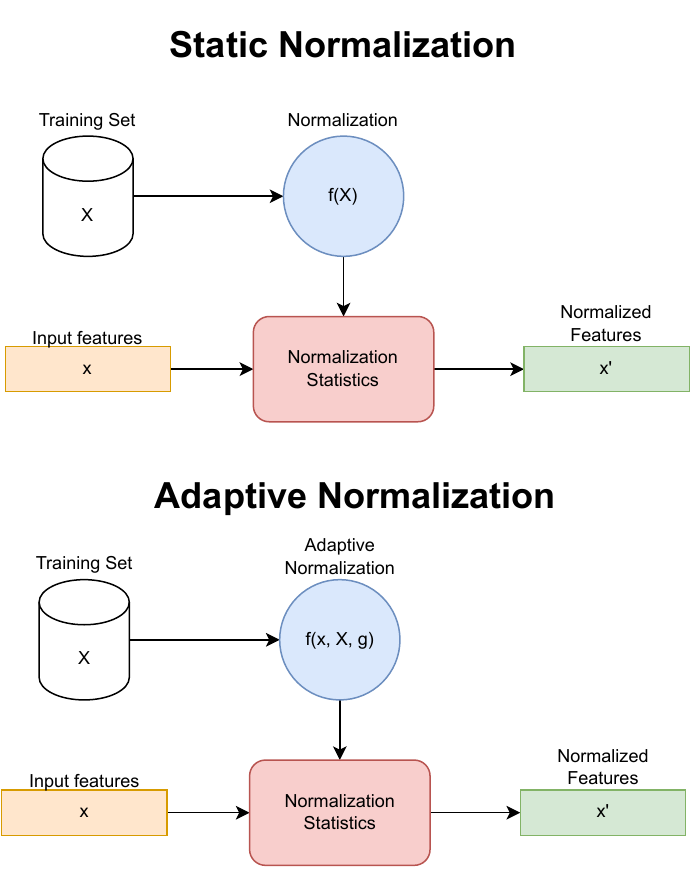}
    \caption{Comparison between static normalization approaches to the adaptive normalization method in \cite{passalis2021forecasting}, which takes into account not only the statistics of the dataset $\mathcal{X}$, but also the distribution from which the current input sample $\mathbf{x}$ is drawn.}
    \label{fig:normalization}
\end{figure}

Commonly used normalization methods, such as min-max normalization or z-score normalization, work best when the input data is stationary. Since financial timeseries can be highly volatile and non-stationary, such approaches are rendered ineffective and can in fact hinder the training process of DL models. To this end, in \cite{passalis2021forecasting}, an adaptive input normalization layer was proposed that can learn to identify the distribution from which the input data were generated and then apply the most appropriate normalization scheme. Static normalization compared against adaptive normalization is illustrated in Figure~\ref{fig:normalization}.

Let $x=\{ \mathbf{x}_1, \mathbf{x}_2, \dots, \mathbf{x}_L \}$, where $\mathbf{x}_i \in \mathbb{R}^d$ a window of a timeseries composed of $L$ measurements, each of them a $d$-dimensional vector in the multivariate case. Assuming a Gaussian distribution for each measurement vector $\mathbf{x}]_l$, they could be effectively normalized as:

\begin{equation}
    [ \mathbf{x}' ]_l = ([\mathbf{x}]_l - \mu_l) / \sigma_l
\end{equation}
where $\mu_l$ and $\sigma_l$ denote the mean and standard deviation for the $l$-th measurement, or variable. However, when the assumption of a Gaussian distribution is relaxed, computing a single mean and standard deviation can cause more harm than good, by applying the same normalization to vastly different parts of the timeseries. A more appropriate alternative is to assume that the measurements were generated by a Gaussian Mixture Model described by the density function:

\begin{equation}
    p(\mathbf{x}) = \sum_{i=1}^{N} \phi_i \mathcal{N}(\boldsymbol{\mu}_i, \boldsymbol{\Sigma}_i)
\end{equation}
where $\phi_i$ is the weight of the $i$-th Gaussian with mean $\boldsymbol{\mu}_i$ and covariance $\boldsymbol{\Sigma}_i$. In this case, we can normalize each window of the timeseries by attempting first to identify the parameters of the Gaussian with which they were generated and then using these parameters to apply normalization. The problem is thus transformed into one of training a neural network to \emph{learn} how the measurements $\mathbf{x}_i$ of a temporal window should be normalized. The learned, adaptive normalization follows the form:

\begin{equation}
\label{eq:norm}
    \mathbf{x}' = (\mathbf{x} - \boldsymbol{\alpha}(x)) \odot \boldsymbol{\beta}(x)
\end{equation}
where the functions $\boldsymbol{\alpha}(x))$ and $\boldsymbol{\beta}(x))$ are not fixed, but depend on the distribution in which the current input sample belongs, and their parameters are learned. The first step of this method is to compute a summary representation by averaging the $L$ measurements in a window:

\begin{equation}
    \mathbf{s}_\alpha (x) = \frac{1}{L} \sum_{i=1}^{L} \in \mathbb{R}^{d}
\end{equation}
which is then used to identify the distribution that generated the current sample. The shifting function $\boldsymbol{\alpha}(x) \in \mathbb{R}^{d}$ is defined as:

\begin{equation}
    \boldsymbol{\alpha}(x) = \mathbf{W}_\alpha\mathbf{s}_\alpha(x) + \mathbf{b}_\alpha
\end{equation}
where $\mathbf{W}_\alpha \in \mathbb{R}^{d \times d}$ and $\mathbf{b}_\alpha \in \mathbb{R}^{d}$ are the weights and bias responsible for transforming the measurements across each dimension. Not that, in practice, this forms allows this function to be implemented as a linear layer, allowing it to be prepended to any neural network and trained end-to-end. After applying this adaptive shift to the input data, it must be appropriately scaled using the scaling function $\boldsymbol{\beta}(x)$. First, an updated summary representation $\mathbf{s}_\beta$ is obtained as:

\begin{equation}
    \mathbf{s}_\beta = \sqrt{\frac{1}{L} \sum_{i=1}^{L} (\mathbf{x}_i - \boldsymbol{a}(x))^2} \in \mathbb{R}^{d}
\end{equation}
corresponding to the standard deviation of the shifted input features.  Then, the scaling function $\boldsymbol{\beta}(x) \in \mathbb{R}^{d}$ can be defined as:

\begin{equation}
    \boldsymbol{\beta}(x) = (\mathbf{W}_\beta\mathbf{s}_\beta + \mathbf{b}_\beta)^{-1}
\end{equation}
where the values of $(\mathbf{W}_\beta\mathbf{s}_\beta + \mathbf{b}_\beta)$ are inverted element-wise, and $\mathbf{W}_\beta \in \mathbb{R}^{d \times d}$ and $\mathbf{b}_\beta \in \mathbb{R}^{d}$ are the weights and bias of the scaling layer.

Having computed both $\boldsymbol{\alpha}(x)$ and $\boldsymbol{\beta}(x)$ the normalized representation $\mathbf{x}'$ can be obtained using Eq. \ref{eq:norm}. Finally, an attention-like gating mechanism is employed to further suppress irrelevant values and noise. A summary $\mathbf{s}_\gamma \in \mathbb{R}^{d}$ is computed for this task as:

\begin{equation}
    \mathbf{s}_\gamma = \frac{1}{L} \sum_{i=1}^{L} \mathbf{x}'_i 
\end{equation}
which is then used to compute the gating function $\boldsymbol{\gamma}(x) \in \mathbb{R}^{d}$ as:

\begin{equation}
    \boldsymbol{\gamma}(x) = \text{sigmoid}(\mathbf{W}_\gamma \mathbf{s}_\gamma(x) + \mathbf{b}_\gamma)
\end{equation}
where $\mathbf{W}_\gamma \in \mathbb{R}^{d\times d}$ and $\mathbf{b}_\gamma \in \mathbb{R}^{d}$ are the weight and bias of the gating layer respectively. Finally, the data is normalized as:

\begin{equation}
    \label{eq:dain}
    \mathbf{x}'' = ((\mathbf{x} - \boldsymbol{\alpha}(x)) \odot \boldsymbol{\beta}(x)) \odot \boldsymbol{\gamma}(x).
\end{equation}

The proposed method was evaluated on top of multiple neural network architectures on the FI-2010 dataset \cite{ntakaris2018benchmark}, as well as on the popular cryptocurrencies BitCoin and Etherium. The results for the FI-2010 dataset are summarized in Table~\ref{tab:normalization}, where it is evident that the proposed method outperforms all of the compared normalization methods, including an adaptive method (DAIN) lacking the gating mechanism which further reduces noise and irrelevant information in the data. The results are reported in terms of F1-score and Cohen's $\kappa$ score, using a prediction horizon of 10 timesteps. Similar results and conclusions are drawn for a longer prediction horizon as well as for the two evaluated cryptocurrencies.

\begin{table}[hb!]
    \caption{FI-2010 evaluation of the RDAIN method \cite{passalis2021forecasting}: evaluating the performance of different normalization methods for the first prediction horizon (next 10 timesteps).}
    \centering
    \begin{tabular}{llll}
        \toprule
         \textbf{Model} & \textbf{Method} & \textbf{F1-score} & $\boldsymbol{\kappa}$ \textbf{score} \\ \midrule
         \textbf{MLP} & Standardization & $56.79 \pm 0.47$ & $0.3536 \pm 0.0071$ \\
         \textbf{MLP} & Sample Average & $56.21 \pm 1.80$ & $0.3426 \pm 0.0235$ \\
         \textbf{MLP} & Sample Standardization & $63.86 \pm 1.31$ & $0.4468 \pm 0.0188$ \\
         \textbf{MLP} & Batch Normalization & $56.26 \pm 0.51$ & $0.3455 \pm 0.0085$ \\
         \textbf{MLP} & Instance Normalization & $60.77 \pm 0.92$ & $0.4006 \pm 0.0114$ \\
         \textbf{MLP} & DAIN & $69.50 \pm 0.35$ & $0.5346 \pm 0.0051$ \\
         \textbf{MLP} & Proposed & $\underline{69.94 \pm 0.55}$ & $\underline{0.5413 \pm 0.0086}$ \\[1em]

         \textbf{CNN} & Standardization & $57.27 \pm 1.12$ & $0.3563 \pm 0.0183$ \\
         \textbf{CNN} & Sample Average & $58.41 \pm 0.63$ & $0.3663 \pm 0.0101$ \\
         \textbf{CNN} & Sample Standardization & $60.44 \pm 1.02$ & $0.3972 \pm 0.0167$ \\
         \textbf{CNN} & Batch Normalization & $55.93 \pm 1.08$ & $0.3372 \pm 0.0182$ \\
         \textbf{CNN} & Instance Normalization & $60.17 \pm 0.69$ & $0.3917 \pm 0.0120$ \\
         \textbf{CNN} & DAIN & $48.87 \pm 3.07$ & $0.2488 \pm 0.0372$ \\
         \textbf{CNN} & Proposed & $\underline{66.77 \pm 0.48}$ & $\underline{0.4926 \pm 0.0086}$ \\[1em]

         \textbf{RNN} & Standardization & $55.48 \pm 1.29$ & $0.3285 \pm 0.0207$ \\
         \textbf{RNN} & Sample Average & $54.24 \pm 1.38$ & $0.3030 \pm 0.0221$ \\
         \textbf{RNN} & Sample Standardization & $59.62 \pm 0.91$ & $0.3850 \pm 0.0151$ \\
         \textbf{RNN} & Batch Normalization & $54.85 \pm 1.32$ & $0.3649 \pm 0.0135$ \\
         \textbf{RNN} & Instance Normalization & $58.33 \pm 0.84$ & $0.3649 \pm 0.0135$ \\
         \textbf{RNN} & DAIN & $67.86 \pm 0.52$ & $0.5102 \pm 0.0085$ \\
         \textbf{RNN} & Proposed & $\underline{67.90 \pm 0.80}$ & $\underline{0.5106 \pm 0.0122}$ \\ \bottomrule
    \end{tabular}
    
    \label{tab:normalization}
\end{table}

In \cite{nalmpantis2021deep}, a group-based adaptive input normalization method was proposed. The input sample is first split into groups before computing the shifting and scaling functions. Each of these are applied to the corresponding grouped input and the normalized output is flattened before being fed to the rest of the network. The gating method added in \cite{passalis2021forecasting} however adds robustness to the normalization and deals with irrelevant values, as shown also in Table~\ref{tab:normalization}. 

To address the high-dimensionality and volatility of financial timeseries data, a Bag-of-Features (BoF) based approach was proposed in \cite{passalis2020temporal}. Typically in BoF, a stationary codebook is created by either clustering or selection in the original input sample space. Instead, a learnable codebook that can be embedded into any neural network model was proposed. The soft-assignment BoF histogram-like representations learn to retain important information while discarding noise and irrelevant values.

\subsection{Addressing the Noisy Labels of Supervised Learning Agents}

With input data normalization addressed, several challenges still remain when handling financial data with DL models. Using supervised learning, label quantization and noisy labels are perhaps the most difficult to overcome. Using reinforcement learning is one approach to avoid creating handcrafted labels, although it comes with its own setback which are addressed in Section~\ref{sec:reinforcement} while this part of the paper focuses solely on supervised learning.

The method proposed in \cite{tsantekidis2021transferring} deals with the problems associated with price quantization by attempting to solve the problem from a different perspective: learning to mimic trading strategies without needing to know the strategy itself. In other words, the DL models are trained to imitate real or artificial trading agents' strategies, discarding the need for price quantization altogether. The DL models learn to generalize this strategy to unseen samples with great accuracy, meaning that they have effectively learned to replicate the decisions of the teacher agents. One additional advantage in this case is that multiple strategies can be combined effective in the form of a DL model ensemble to create an agent that incorporates knowledge from different sources.

In \cite{floratos2022online}, an ensemble of DL models is used as a method to \emph{soften} the labels that a student DL network learns similarly to how Knowledge Distillation (KD) methods work. Price quantization is performed in this case and its drawbacks are highlighted on FOREX data, where the labels are extremely noisy. Going one step further more traditional KD, the teachers ensemble and the student network are trained in a single step, effectively reducing the number of hyperparameters that need to be tuned for a system like this to be trained. 

Formally, let $\mathbf{y}_t = f_T(\mathbf{x})$ and $\mathbf{y}_S = f_S(\mathbf{x})$ denote the functions of a teacher network and a student network respectively. Knowledge distillation is a method used to facilitate the training of student models under the supervision of a strong teacher network. The soft outputs produced by the teacher are more informative about the data than standard hard, binary labels as they are enriched with information about hidden data similarities as defined by the teacher network's weights. Therefore, KD can be seen as a type of learned label smoothing, where to ensure the smoothness of the labels the softmax function with temperature can be used:

\begin{equation}
    p_i = \frac{\text{exp}(z_i / T)}{\sum_{i} \text{exp}(z_i/T)}
\end{equation}
where $z_i$ are the teacher's predicted logits, $T$ denotes the temperature parameter and $p_i$ are the produced smooth labels for all $i=1, ..., C$ classes. As $T$ increases, the probability distribution produced by the softmax becomes softer, magnifying the data similarities that were probably concealed under negligible probabilities. Although typically a combination of the hard and smooth labels is used to train a student model, in this work only the softened labels were used as the hard labels were deemed far too noisy to be useful. 

\begin{figure}
    \centering
    \includegraphics[width=0.8\linewidth]{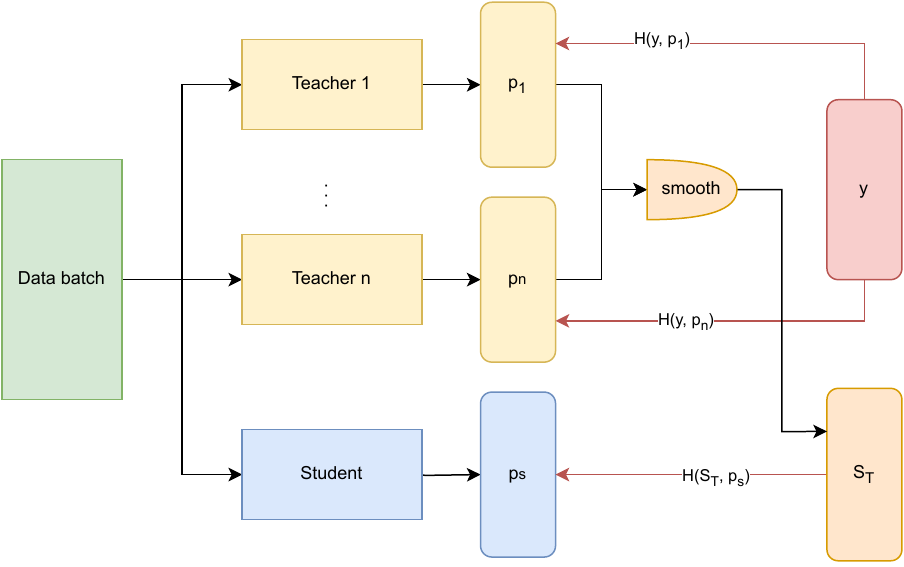}
    \caption{Overview of the Online Self Knowledge Distillation  method proposed in \cite{floratos2022online}.}
    \label{fig:oskd}
\end{figure}

The proposed Online KD (OKD) method is based on the same principles but trains an ensemble of teachers, instead of a single teacher, and the student networks in a single phase. Specifically the teachers are trained using the hard labels and the cross-entropy loss function:

\begin{equation}
    H(\mathbf{y}, \mathbf{p}_t) = \sum_{i=1}^{C} y(i) \text{log}(p_t(i))
\end{equation}
where $\mathbf{y}$ are the hard ground truth class labels, $C$ is the number of classes (i.e., 3 for price trend prediction), and $\mathbf{p}_t$ is the output probability distribution of teacher $t$. Then, using $N$ teachers, the soft labels are formulated as:

\begin{equation}
    \mathbf{S}_T = \frac{1}{N} \sum_{i=1}^{N} \mathbf{p}_t
\end{equation}
and finally the teacher network is trained using the cross-entropy objective using the soft labels as the ground truth:
\begin{equation}
    H(\mathbf{S}_T, \mathbf{p}_s) = - \sum_{i=1}^{C} S_T(i) \text{log}(p_s(i)).
\end{equation}
All of these losses are computed in parallel using the same input batch for each training step. The final objective of this OKD system is formulated as:
\begin{equation}
    \mathcal{L}(\{\mathbf{p}_t\}_{t=1}^{N}, \mathbf{p}_s, \mathbf{y}; \{\theta_t\}_{t=1}^{N}, \theta_s) = H(\mathbf{S}_T, \mathbf{p}_s) + \lambda \sum_{t=1}^{N} H(\mathbf{y}, \mathbf{p}_t)
\end{equation}
where $\lambda$ is a factor that weighs the contribution of the teachers' losses. The proposed framework is graphically illustrated in Figure~\ref{fig:oskd} Further advancing the idea of self distillation (SD), the introduction of student self distillation to the process was proposed. With SD, the output probability distribution of the network itself are used in the formation of the final soft targets that will be utilized for its training.

The experimental study is summarized in Table~\ref{tab:supervised} in terms of PnL and accuracy on the test set of a dataset composed of FOREX trading data of 37 different currencies, such as EUR/USD, CHF/JPY, GBP/CAD, USD/NOK, in the period of 2010-2019. The proposed (*) online self teacher-student distillation method achieved the best results among all experiments.

\begin{table}[]
    \caption{Test results comparison amongst different KD methods in supervised learning.}
    \label{tab:supervised}
    \centering
    \begin{tabular}{lll}
    \toprule
         & \textbf{PnL} & \textbf{Accuracy} \\ \midrule
        \textbf{Baseline} & $0.128\%$ & $40.091\%$ \\
        \textbf{KD} & $0.757\%$ & $42.453\%$ \\ 
        \textbf{OKD} & $0.687\%$ & $42.328$ \\
        \textbf{*Teachers SD} & $\underline{2.322\%}$ & $43.092\%$ \\
        \textbf{*Teachers-Student} SD & $2.268\%$ & $\underline{43.109\%}$ \\ \bottomrule
    \end{tabular}
    
\end{table}

\subsection{Addressing the Instability of Reinforcement Learning Agents}
\label{sec:reinforcement}


RL consists of two main components, the environment and an agent interacting with said environment via predefined actions. In the case of financial trading, the environment consists of a mechanism that can simulate a trading desk when given past market data. The environment can also provide the agent with observations regarding the market. The observation of the environment along with the current position of the agent is the \emph{state} of the environment, and is denoted as $s_t$ for the state of time-step $t$. The agent in this context is given three choices at each time-step: buy, sell or exit. The position of the agent at time-step $t$ is denoted as $\delta_t$ and can take the values $\{-1, 0, 1\}$ for the positions short (sell), stay (exit) out of the market, and long (buy), respectively. Depending on the agent's actions, a reward $r_t$ is received. This reward can typically correspondong to the PnL that the agent's action incurs in the market:

\begin{equation}
    r_t^{(PnL)} = \begin{cases}
        z_t & \text{if agent is long} \\
        -z_t & \text{if agent is short} \\
        0 & \text{if agent has no position}
    \end{cases}
\end{equation}
where $z_t$ is the percentage return (i.e., percentage change of the close price) at time-step $t$, simplifying the reward function as $r_t^{(PnL)} = z_t \cdot \delta_t$. Furthermore, in a real trading environment, the agent also pays a cost to change position, called \emph{commission}. This can be incorporated into the reward of the agent as $r_t^{(Fee)} = -c \cdot |\delta_t - \delta_{t-1}|$ where $c$ is the commission cost.

Although RL mitigates the difficulties of selecting suitable thresholds for supervised labels it still suffers from the noisy PnL rewards. This problem is further exaggerated when training an agent on multiple financial assets of highly varying values, even though being able to train across multiple asset pairs and track reoccurring patterns could intuitively increase the performance and stability of the RL training procedure. 

A concept that has assisted in solving hard problems with RL is reward shaping, which attempts to smoothly distribute the rewards along each training epoch. In \cite{tsantekidis2020price}, a reward shaping method that provides more consistent rewards to the agent during its initial interactions with the environment was employed to mitigate the large variance of the rewards caused by the noisy nature of the PnL-based rewards, significantly improving the profitability of the learned
trading policies. 

Specifically, a price trailing objective is added to the agent's learning objective. The agent is assigned its own price value $p_a(t)$ which it can control by either upward or downward increments. The agent's price is compared against the target price $p_\tau(t)$ which acts as the mid-point of the trajectory which the agent is rewarded the most to follow, e.g., set to the close price. The proposed trailing reward is defined as:
\begin{equation}
    \label{eq:trail}
    r_t^{(Trail)} = 1 - \frac{|p_a(t) - p_c(t)|}{m\cdot p_c(t)}
\end{equation}
where $m$ is a margin fraction, which controls how close to the target price the agent's price should stay. Finally, the PnL, Fee and Trail rewards are normalized using the percentage returns' statistics to keep them within more or less the same range. 

The method was evaluated on top of two DRL methods, namely Double Deep Q-Learning (DDQN) and Proximal Policy Optimization (PPO). The methods were evaluated on a financial dataset that contains 28 different instrument combinations with currencies such as Euro, Dollar, 
 British pounds and Canadian dollar among others. The dataset contains minute price candles starting from 2009 to mid-2018. According to the experimental study in this work, optimizing RL agents using PPO lead to significantly improved performance over the DDQN counterparts, although the proposed reward shaping policy proved helpful in both cases compared to the baseline PnL and Fee rewards. The results using PPO are summarised in Table~\ref{tab:reinforcement} in terms of PnL, Sharpe ratio and max drawdown, which is calculated as the maximum percentage difference of the highest peak in profits to the lowest following drop. 

\begin{table}[hb!]
    \caption{Metric results comparing DRL agents trained using Proximal Policy Optimization without and with the proposed price trailing method.}
    \label{tab:reinforcement}
    \centering
    \begin{tabular}{llll}
    \toprule
         & \textbf{Without trailing} & \textbf{With trailing (PPO)} \\ \midrule
        \textbf{PnL} & $23.5\% \pm 2.0\%$ &  $\underline{28.6\% \pm 2.4\%}$ \\
        \textbf{Sharpe} ratio & $3.59 \pm 0.37$ & $\underline{4.20 \pm 0.43}$ \\
        \textbf{Drawdown} & $1.6\% \pm 0.6\%$ & $\underline{1.7\% \pm 0.5\%}$ \\
        \bottomrule
    \end{tabular}
\end{table}

A different approach to stabilizing the training process of DRL agents is to use KD \cite{tsantekidis2021diversity}, in a similar fashion to how it was described in the supervised learning setting. In order to diversify the knowledge of the teacher ensemble, each network in it is trained on a different subset of the RL environment. The teachers' actions are softened and the student network is trained to mimic their soft versions using the Kullback-Leibler divergence. 

In \cite{rodinos2023sharpe}, the Sharpe ratio is incorporated into the reward of a PPO-trained RL agent with the objective to improve the overall performance of the portfolio, by mitigating the risk that occurs in the agent’s decisions. The Sharpe ratio, as described in Section~\ref{sec:metrics}, takes into account the standard deviation of the returns, and optimizing the sharpe ratio means aiming for a stable, yet small, standard deviation of returns.

\subsection{Leveraging Sentiment Information in DL-based Trading Agents}
%
%
%

Despite the success achieved by the methods described in the previous sections, the majority of them primarily rely on price-related data. This data ranges from coarse information, such as minute/hourly/daily OLHC candles, to more detailed information from the limit order book, including volume and price of requested limit and market orders. However, this approach contrasts with how most human traders make decisions when buying or selling assets. Human traders typically consider a wide range of factors and information sources beyond just price and order book data. These factors can include sentiment expressed in various environments such as news and social media, as well as prior knowledge and forecasts, which can significantly impact the price of financial assets~\cite{mai2018does, burnie2019social, guegan2021does}.

Recent studies have shown that approaches utilizing sentiment-related data, often obtained from online sources, can yield promising results and frequently outperform methods relying solely on price information. For example, in~\cite{passalis2021learning} and~\cite{passalis2022multisource}, the authors explore the use of sentiment information extracted from various online sources, such as news articles, in training deep learning agents for financial trading. The authors evaluated different deep learning models as sentiment extractors and introduced a multi-source sentiment fusion approach, demonstrating that sentiment information can be a stronger predictor for Bitcoin trading compared to the actual price time-series. Similar results have been also obtained on generic cryptocurrency trading~\cite{avramelou2023w}, validating the importance of employing sentiment sources in such markets.

In particular, in \cite{avramelou2023w}, a dataset containing sentiment information about cryptocurrencies was made publicly available and evaluated in a DRL setting by itself and in combination with price related information. Sentiment data was extracted using text that was collected from various online sources, such as online articles and social media platforms. Text examples have been collected based on keywords related to cryptocurrency and sentiment analysis was performed using FinBERT \cite{araci2019finbert} to predict a 3-label (positive, negative, neutral) sentiment of each sample of text. The final sentiment score for each document $\mathbf{d}$ was calculated as the difference of positive sentiment score $o_{p}(\mathbf{d})$ minus negative sentiment score $o_{n}(\mathbf{d}$:
\begin{equation}
f(\mathbf{d})  = o_{p}(\mathbf{d}) - o_{n}(\mathbf{d}).
\end{equation}
This indicates that positive texts should have an overall score around 1, negative texts should have an overall score close to -1, and neutral texts have a sentiment score around 0. The collected dataset is published in Zenodo~\cite{loukia_avramelou_2023_7684410} and is publicly available at 
 \url{https://doi.org/10.5281/zenodo.7684410}

The dataset spans from from 17-08-2017 to 12-02-2022 and sentiment from all collected sources is aggregated on a minute basis. For price related information, OHLC price data from 14 cyrptocurrency-USDT pairs was used. The DL model used was an LSTM trained using the PPO method of reinforcement learning. The results, summarized in Table~\ref{tab:sentiment}, indicate that even by itself sentiment information is useful to DRL agents and can be used to train agents that make profitable actions. However, the results using sentiment only pale in comparison to using price related information only. When the two modes are combined, this leads to the most profitable and stable agents in the training and test sets. This indicates that sentiment information can effectively be combined with price related information and its potential can indeed be harvested successfully by DRL agents.

\begin{table}[]
\centering
\caption{Mean train and test PnL after DRL agent training with price-only, sentiment-only and a combination of both types of features.}
\small
\label{tab:sentiment}
\begin{tabular}{lll}
\toprule
                       & \textbf{Test PnL }             & \textbf{Train PnL } \\ \midrule
\textbf{Sentiment-only} & 6.69 $\pm$ 6.43 & 116.92 $\pm$ 20.58 \\ 
\textbf{Price only}             & 9.68 $\pm$  4.93             & 149.08 $\pm$ 5.37 \\ 
\textbf{Price + Sentiment} & \underline{10.55 $\pm$ 1.16} & \underline{156.68 $\pm$ 5.38} \\ \bottomrule
\end{tabular}
\end{table}

However, in general, incorporating such approaches into production systems can be challenging due to the additional requirement of developing data collection pipelines. Gathering sentiment-related information typically involves collecting data from various online sources and social media platforms, followed by preprocessing to extract relevant information usable by deep learning models. Consequently, the development, setup, and maintenance of these pipelines can pose significant obstacles to the integration of such approaches into production systems.  To overcome these limits, methods that use sentiment information only during training, such as~\cite{panagiotatos2022sentiment}, which relies on cross-modal distillation, have been proposed. Such approaches allow for improving the performance of forecasting models while relaxing the need to collect sentiment in real time and maintain the corresponding pipelines.


Furthermore, the aforementioned method either focus on one asset or, when studying multiple assets, they make use of aggregated sentiment information of all the assets studied or information related to the \emph{type} of asset used, e.g., using generic cryptocurrency keywords when studying cryptocurrencies. This discards important information regarding the intricacies of each asset studied, for example one would argue that when studying cryptocurrencies, the extremely popular BitCoin should be given a different weight than other, unpopular assets, even though the latter shouldn't be entirely ignored as their price information might contain informative patterns and additional, useful training samples. This highlights the need for fine-grained sentiment information, in the sense of information per asset instead of generic sentiment towards the type of traded assets.




\section{Conclusion}
\label{sec:conclusions}

Algorithmic financial trading has been gaining more and more in popularity and deploying Deep Learning models as trading agents has been a promising approach with impressive results in recent years. However, financial data comes with several challenges when combined with DL models. In this paper, we have identified several challenges that affect DL models when applied to financial data and discussed various methods to alleviate the negative effects of these challenges on the models. First, financial timeseries can be highly non-stationary, as well as very high-dimensional, and exhibit wide variety and great volatility. Methods for adaptive data normalization and effective feature extraction were discussed to amend this. Second, another great challenge in dealing with financial data arises from the need of price quantization in the supervised learning case of price trend prediction. This quantization can lead to data imbalance and, more importantly, noisy labels that can severely hinder the convergence of a DL model. Methods based on KD and label smoothing using ensembles of teacher networks were discussed as possible solutions.

Although reinforcement learning techniques can completely overcome this problem as they do not require label quantization, they tend to exhibit instability during training and deployment that needs to be addressed. Reward shaping methods which keep rewards steady and withing certain limits were proposed to address the instability. Furthermore, KD methods were shown to be effective in the case of DRL agents as well in reducing instability. Finally, when combining sentiment information with price-related features to improve the trading agents' performance, several challenges were identified and solutions proposed. The quality of sentiment information sources and DL extractors was addressed. The usefulness of this information pertaining to the financial trading is a more challenging and open problem in this task. Both in supervised and reinforcement learning settings, the incorporation of sentiment information as another modality was studied and its usefulness in the financial trading problem was established.

Despite advances, these challenges remain open and are the driving forces behind recent research in the subject. Recent advances in DL areas such as self-supervised learning, generative models, explainable AI and neuromorphic models, among other directions, are promising research directions for the future of algorithmic financial trading. 

\section*{Acknowledgments}
This work is co-ﬁnanced by the European Regional Development Fund of the European Union and Greek national funds through the Operational Program Competitiveness, Entrepreneurship and Innovation, under the call RESEARCH - CREATE - INNOVATE (project code: T2EDK-02094).

\bibliographystyle{unsrt}  
\bibliography{references}

\end{document}